\documentclass[twocolumn,preprintnumbers,amsmath,amssymb]{revtex4}
\usepackage{graphicx}% Include figure files
\usepackage{dcolumn}% Align table columns on decimal point
\usepackage{bm}% bold math

\DeclareMathOperator{\erf}{erf}
\DeclareMathOperator{\erfc}{erfc}
\newcommand{\bra}[1]{\ensuremath{\langle #1 \vert}}
\newcommand{\ket}[1]{\ensuremath{\vert #1  \rangle}}

\renewcommand{\b}[1]{\ensuremath{\mathbf{#1}}}
\newcommand{\coul}{\ensuremath{\text{coul}}}
\newcommand{\lr}{\ensuremath{\text{lr}}}
\newcommand{\sr}{\ensuremath{\text{sr}}}
\newcommand{\unif}{\ensuremath{\text{unif}}}

\newcommand{\LDA}{\ensuremath{\text{LDA}}}

\renewcommand{\H}{\ensuremath{\text{H}}}

\begin{document}

\title{Local density approximation for long-range or for short-range energy functionals?}

\author{Julien Toulouse}
\email{toulouse@lct.jussieu.fr}
\author{Andreas Savin}
\email{savin@lct.jussieu.fr}
\affiliation{
Laboratoire de Chimie Th\'eorique, CNRS et Universit\'e Pierre et Marie Curie,\\
4 place Jussieu, 75252 Paris, France.
}

\date{\today}
             
\begin{abstract}
Density functional methods were developed, 
in which the Coulomb electron-electron interaction is split into a 
long- and a short-range part. 
In such methods, one term is calculated using traditional density functional approximations, like the local density approximation.
The present paper tries to shed some light upon the best way to do it
by comparing the accuracy of the local density approximation with accurate results for the He atom.

\end{abstract}

\maketitle

\section{Introduction}
\label{sec:intro}

In recent years, there has been a growing interest in approaches of density functional theory (DFT)~\cite{HohKoh-PR-64} based on a long-range/short-range decomposition of the Coulomb electron-electron interaction. The idea is to use different, appropriate approximations for the long-range and the short-range contributions to the exchange and/or correlation energy density functionals of the Kohn-Sham (KS) scheme~\cite{KohSha-PR-65}. 

Various schemes combining a (semi)local short-range exchange energy functional approximation with an ``exact'' long-range exchange energy expression have been proposed (see, e.g., Refs.~\onlinecite{IikTsuYanHir-JCP-01,KamTsuHir-JCP-02,TawTsuYanYanHir-JCP-04,YanTewHan-CPL-04,BaeNeu-PRL-05,GerAng-CPL-05}), which allow to impose the exact Coulombic asymptotic behavior $1/r$ of the exchange interaction. 
This turned out to be important for charge transfer, van der Waals interactions, etc.
But opposite approaches combining a (semi)local long-range exchange functional approximation with an ``exact'' (or ``hybrid'') short-range exchange have also been used (see, e.g., Refs.~\onlinecite{BylKle-PRB-90,SeiGorVogMajLev-PRB-96,HeyScuErn-JCP-03,HeyScu-JCP-04,HeyScu-JCP-04b}), which allow to introduce exact exchange in solid-state calculations without the computationally-demanding exact long-range contribution. 
There is also a physical reason for doing it: long-range correlations are 
not well treated by (semi)local density approximations. 
Treating only the exchange exactly destroys the balance of errors, which is important for metals, gaps, etc.
Contemplating this two opposite approaches, one can ask: \textit{is it preferable to use (semi)local density functional approximations for the long-range or for the short-range contribution to the exchange energy?}

Nozi\`eres and Pines~\cite{NozPin-PR-58} have first used the idea of decomposing the correlation energy of the uniform electron gas into long-range and short-range contributions to facilitate its calculation. In the context of DFT, a few schemes combining a (semi)local short-range density functional approximation with a long-range correlation energy calculated by other means have been proposed (see, e.g., Refs.~\onlinecite{KohHan-JJJ-XX,KohMeiMak-PRL-98,TouColSav-PRA-04,PedJen-JJJ-XX,AngGerSavTou-PRA-05}) to deals with for example near-degeneracy or long-range van der Waals interactions. Indeed, it is well known that  (semi)local density functional approximations are appropriate for the short-range contribution (see, e.g., Refs.~\onlinecite{LanPer-PRB-77,BurPer-IJQC-95}). However, for a given decomposition of the Coulomb interaction, there are at least two possible definitions for a short-range correlation functional. It can be defined either as the difference between the Coulombic correlation functional and a long-range correlation functional associated to the long-range part of interaction (as in Ref.~\onlinecite{TouColSav-PRA-04}), or directly from the short-range part of the interaction (as in Ref.~\onlinecite{KohMeiMak-PRL-98}). Therefore, one can wonder: \textit{what is the preferable definition for a short-range correlation functional as regards as the accuracy of (semi)local density functional approximations?}

This work sheds some light on these two questions. Taking the example of the He atom, we test the accuracy of the local density approximation (LDA) to the long-range and short-range contributions to the exchange and correlation energies for a given decomposition of the interaction. Atomic units (a.u.) are used throughout this paper.

\section{Long-range and short-range density functionals}

We define in this section the long-range and short-range density functionals that we consider in this work.

Let's define first a general exchange functional for an arbitrary electron-electron interaction $w_{ee}(r)$
\begin{eqnarray}
E_{x}[n;w_{ee}] = \bra{\Phi[n]} \hat{W}_{ee} \ket{\Phi[n]} - E_{\H}[n;w_{ee}],
\end{eqnarray}
where $\hat{W}_{ee} = (1/2) \iint \hat{n}_2(\b{r}_1,\b{r}_2) w_{ee}(r_{12}) d\b{r}_1 d\b{r}_2$ is the interaction operator, expressed with the pair-density operator $\hat{n}_2(\b{r}_1,\b{r}_2)$, $\Phi[n]$ is the KS determinant and $E_{\H}[n;w_{ee}] = (1/2) \iint n(\b{r}_1) n(\b{r}_2) w_{ee}(r_{12}) d\b{r}_1 d\b{r}_2$ is the Hartree functional. Likewise, the general correlation functional associated to $w_{ee}(r)$ writes
\begin{eqnarray}
E_{c}[n;w_{ee}] = \min_{\Psi \to n} \bra{\Psi} \hat{T} + \hat{W}_{ee} \ket{\Psi} - \bra{\Phi[n]} \hat{T} + \hat{W}_{ee} \ket{\Phi[n]},
\nonumber\\
\end{eqnarray}
where the constrained-search formalism~\cite{Lev-PNAS-79} has been used.

We use the following decomposition of the Coulomb electron-electron interaction $w_{ee}^{\coul}(r) = 1/r$
\begin{eqnarray}
w_{ee}^{\coul}(r) = w_{ee}^{\lr,\mu}(r) + w_{ee}^{\sr,\mu}(r),
\end{eqnarray}
where $w_{ee}^{\lr,\mu}(r)=\erf(\mu r)/r$ is a long-range interaction and $w_{ee}^{\sr,\mu}(r)=\erfc(\mu r)/r$ is its short-range complement. The parameter $\mu$ which controls the range of the decomposition varies between $0$ and $\infty$. For $\mu=0$, the long-range interaction vanishes, $w_{ee}^{\lr,\mu=0}=0$, and the short-range interaction reduces to the Coulomb interaction $w_{ee}^{\sr,\mu=0}=w_{ee}^{\coul}$. In the limit $\mu \to \infty$, the long-range interaction reduces to the Coulomb interaction $w_{ee}^{\lr,\mu \to \infty}=w_{ee}^{\coul}$, and the short-range interaction vanishes, $w_{ee}^{\sr,\mu \to \infty}=0$.

The Coulombic exchange functional $E_{x}^{\coul}[n] = E_{x}[n;w_{ee}^{\coul}]$ is consequently decomposed as
\begin{eqnarray}
E_{x}^{\coul}[n] = E_{x}^{\lr,\mu}[n] + E_{x}^{\sr,\mu}[n],
\label{Exdecomp}
\end{eqnarray}
where $E_{x}^{\lr,\mu}[n] = E_{x}[n;w_{ee}^{\lr,\mu}]$ is the long-range exchange functional associated to the interaction $w_{ee}^{\lr,\mu}$, and $E_{x}^{\sr,\mu}[n] = E_{x}[n;w_{ee}^{\sr,\mu}]$ is the short-range exchange functional associated to the interaction $w_{ee}^{\sr,\mu}$.

One can also define a long-range correlation functional associated to the interaction $w_{ee}^{\lr,\mu}$,  $E_{c}^{\lr,\mu}[n] = E_{c}[n;w_{ee}^{\lr,\mu}]$, and a short-range correlation functional associated to the interaction $w_{ee}^{\sr,\mu}$,  $E_{c}^{\sr,\mu}[n] = E_{c}[n;w_{ee}^{\sr,\mu}]$. However, because $E_{c}[n;w_{ee}]$ is not linear with respect to $w_{ee}$, we do not have for the Coulombic correlation functional $E_{c}^{\coul}[n] = E_{c}[n;w_{ee}^{\coul}]$ the same decomposition as in Eq.~(\ref{Exdecomp})
\begin{eqnarray}
E_{c}^{\coul}[n] \neq E_{c}^{\lr,\mu}[n] + E_{c}^{\sr,\mu}[n].
\label{}
\end{eqnarray}
Instead, one can write $E_{c}^{\coul}[n]$ exactly as
\begin{eqnarray}
E_{c}^{\coul}[n] = E_{c}^{\lr,\mu}[n] + E_{c}^{\sr,\mu}[n] + E_{c}^{\lr-\sr,\mu}[n],
\label{Ecdecomp0}
\end{eqnarray}
defining the mixed long-range/short-range correlation functional $E_{c}^{\lr-\sr,\mu}[n]$ which encompasses all the terms stemming from the non-linearity of $E_{c}[n;w_{ee}]$ with respect to $w_{ee}$. By associating the mixed term $E_{c}^{\lr-\sr,\mu}[n]$ to either the long- or the short-range part of the correlation energy, one obtains two possible long-range/short-range decompositions. The first one is
\begin{eqnarray}
E_{c}^{\coul}[n] = E_{c}^{\lr,\mu}[n] + \bar{E}_{c}^{\sr,\mu}[n],
\label{Ecdecomp1}
\end{eqnarray}
defining the new short-range correlation functional $\bar{E}_{c}^{\sr,\mu}[n] = E_{c}^{\sr,\mu}[n] + E_{c}^{\lr-\sr,\mu}[n]$. The second one is
\begin{eqnarray}
E_{c}^{\coul}[n] = \bar{E}_{c}^{\lr,\mu}[n] + E_{c}^{\sr,\mu}[n],
\label{Ecdecomp2}
\end{eqnarray}
defining the new long-range correlation functional $\bar{E}_{c}^{\lr,\mu}[n] = E_{c}^{\lr,\mu}[n] + E_{c}^{\lr-\sr,\mu}[n]$. 

All the long-range functionals vanish for $\mu=0$, $E_{x}^{\lr,\mu=0}=E_{c}^{\lr,\mu=0}=\bar{E}_{c}^{\lr,\mu=0}=0$, and reduce to the Coulombic functionals for $\mu \to \infty$, $E_{x}^{\lr,\mu \to \infty} = E_{x}^{\coul}$ and $E_{c}^{\lr,\mu \to \infty} = \bar{E}_{c}^{\lr,\mu \to \infty} = E_{c}^{\coul}$. Symmetrically, all the short-range functionals reduce to the Coulombic functionals for $\mu=0$, $E_{x}^{\sr,\mu=0} = E_{x}^{\coul}$ and $E_{c}^{\sr,\mu=0} = \bar{E}_{c}^{\sr,\mu=0} = E_{c}^{\coul}$, and vanish for $\mu \to \infty$, $E_{x}^{\sr,\mu \to \infty}=E_{c}^{\sr,\mu \to \infty}=\bar{E}_{c}^{\sr,\mu \to \infty}=0$. The mixed long-range/short-range correlation functional vanishes for both $\mu=0$ and $\mu \to \infty$, $E_{c}^{\lr-\sr,\mu=0}=E_{c}^{\lr-\sr,\mu \to \infty}=0$.

At first sight, the physical meaning of the decomposition of Eq.~(\ref{Ecdecomp0}) may not be obvious. To get more insight on the nature of the terms $E_{c}^{\lr,\mu}$, $E_{c}^{\sr,\mu}$ and $E_{c}^{\lr-\sr,\mu}$, we show now the expressions of these functionals in second-order G\"{o}rling-Levy perturbation theory~\cite{GorLev-PRA-94}. We start from the second-order Coulombic correlation energy
\begin{eqnarray}
E_{c}^{\coul,(2)} = \sum_i \frac{|\bra{\Phi} \hat{W}_{ee}^{\coul} - \hat{V}_{\H x}^{\coul} \ket{\Phi_i} |^2}{E_s - E_{s,i}},
\label{Eccoul2}
\end{eqnarray}
where $\hat{V}_{\H x}^{\coul} = \int \hat{n}(\b{r}) \delta E_{\H x}^{\coul}/\delta n(\b{r}) d\b{r}$ is the Coulombic Hartree-exchange potential operator expressed with density operator $\hat{n}(\b{r})$ and with the Coulombic Hartree-exchange functional $E_{\H x}^{\coul}[n]$, $\Phi$ and $E_s$ are the KS wave function and energy, $\Phi_i$ and $E_{s,i}$ are the excited KS eigenfunctions and eigenvalues. Applying in Eq.~(\ref{Eccoul2}) the long-range/short-range decomposition on $\hat{W}_{ee}^{\coul}$ and $\hat{V}_{\H x}^{\coul}$ and expanding leads to the second-order expressions for $E_{c}^{\lr,\mu}$, $E_{c}^{\sr,\mu}$ and $E_{c}^{\lr-\sr,\mu}$: 

\begin{itemize}
\item The second-order long-range correlation energy writes
\begin{eqnarray}
E_{c}^{\lr,\mu,(2)} = \sum_i \frac{|\bra{\Phi} \hat{W}_{ee}^{\lr,\mu} - \hat{V}_{\H x}^{\lr,\mu} \ket{\Phi_i} |^2}{E_s - E_{s,i}},
\label{}
\end{eqnarray}
where $\hat{V}_{\H x}^{\lr,\mu} = \int \hat{n}(\b{r}) \delta E_{\H x}^{\lr,\mu}/\delta n(\b{r}) d\b{r}$ is the long-range Hartree-exchange potential operator expressed with the long-range Hartree-exchange functional $E_{\H x}^{\lr,\mu}[n]$ defined in the same way as the long-range exchange functional. 

\item Likewise, the second-order short-range correlation energy writes
\begin{eqnarray}
E_{c}^{\sr,\mu,(2)} = \sum_i \frac{|\bra{\Phi} \hat{W}_{ee}^{\sr,\mu} - \hat{V}_{\H x}^{\sr,\mu} \ket{\Phi_i} |^2}{E_s - E_{s,i}},
\label{}
\end{eqnarray}
where $\hat{V}_{\H x}^{\sr,\mu} = \int \hat{n}(\b{r}) \delta E_{\H x}^{\sr,\mu}/\delta n(\b{r}) d\b{r}$ is the short-range Hartree-exchange potential operator expressed with the short-range Hartree-exchange functional $E_{\H x}^{\sr,\mu}[n]$ defined in the same way as the short-range exchange functional. 

\item Finally, the second-order mixed long-range/short-range correlation functional is
\begin{eqnarray}
E_{c}^{\lr-\sr,\mu,(2)} &=&
\nonumber\\ 2
\sum_i \frac{\bra{\Phi} \hat{W}_{ee}^{\lr,\mu} - \hat{V}_{\H x}^{\lr,\mu} \ket{\Phi_i} \bra{\Phi_i} \hat{W}_{ee}^{\sr,\mu} - \hat{V}_{\H x}^{\sr,\mu} \ket{\Phi} }{E_s - E_{s,i}}.
\label{}
\end{eqnarray}

\end{itemize}

\section{Local density approximation}

For a given electron-electron interaction, the local density approximation to the previously introduced exchange and correlation functionals consists in locally transferring the corresponding energy of a uniform electron gas with the same interaction and with density equal to the local value of the inhomogeneous density.

Let us first consider the short-range functionals. The short-range exchange LDA functional associated to $E_{x}^{\sr,\mu}[n]$ writes
\begin{eqnarray}
E_{x,\LDA}^{\sr,\mu}[n] = \int n(\b{r}) \varepsilon_{x,\unif}^{\sr,\mu}(n(\b{r})) d\b{r},
\label{ExsrLDA}
\end{eqnarray}
where $\varepsilon_{x,\unif}^{\sr,\mu}(n)$ is the exchange energy per particle of a uniform electron gas with interaction $w_{ee}^{\sr,\mu}$~\cite{Sav-INC-96,TouSavFla-IJQC-04}. Similarly, the short-range correlation LDA functional for $\bar{E}_{c}^{\sr,\mu}[n]$ is
\begin{eqnarray}
\bar{E}_{c,\LDA}^{\sr,\mu}[n] = \int n(\b{r}) \bar{\varepsilon}_{c,\unif}^{\sr,\mu}(n(\b{r})) d\b{r},
\label{EcsrcompLDA}
\end{eqnarray}
where $\bar{\varepsilon}_{c,\unif}^{\sr,\mu}(n)=\varepsilon_{c,\unif}^{\coul}(n) - \varepsilon_{c,\unif}^{\lr,\mu}(n)$ is obtained as the difference of the Coulomb correlation  energy, $\varepsilon_{c,\unif}^{\coul}(n)$, and of the long-range correlation energy of a uniform electron gas with interaction $w_{ee}^{\lr,\mu}$~\cite{Sav-INC-96,TouSavFla-IJQC-04}. The LDA for $E_{c}^{\sr,\mu}[n]$ writes
\begin{eqnarray}
E_{c,\LDA}^{\sr,\mu}[n] = \int n(\b{r}) \varepsilon_{c,\unif}^{\sr,\mu}(n(\b{r})) d\b{r},
\label{EcsrLDA}
\end{eqnarray}
where $\varepsilon_{c,\unif}^{\sr,\mu}$ is the correlation energy per particle of a uniform electron gas with interaction $w_{ee}^{\sr,\mu}$~\cite{ZecGorMorBac-PRB-04}.

The LDA corresponding to the long-range functionals are obtained by difference to the Coulombic case: $E_{x,\LDA}^{\lr,\mu}[n] = E_{x,\LDA}^{\coul}[n] - E_{x,\LDA}^{\sr,\mu}[n]$, $E_{c,\LDA}^{\lr,\mu}[n] = E_{c,\LDA}^{\coul}[n] - \bar{E}_{c,\LDA}^{\sr,\mu}[n]$ and $\bar{E}_{c,\LDA}^{\lr,\mu}[n] = E_{c,\LDA}^{\coul}[n] - E_{c,\LDA}^{\sr,\mu}[n]$.

\section{Results for the He atom}

We consider the simple example of the He atom. We compare the accuracy of the LDA for the long-range or short-range energies as follows. For each value of mu (selected in $[0,\infty[$ ), we obtain an accurate value of the energy (see Refs.~\onlinecite{PolColLeiStoWerSav-IJQC-03,TouColSav-PRA-04} for details), we calculate the LDA error on the energy, and report in the plots the LDA error (ordinate) for each accurate energy (abscissa). We choose this way of plotting in order to emphasize the importance of the error for a given energy value to be recovered by the approximation: a good functional would yield small errors even for large contributions of the energy. Furthermore, this representation gives a unique scale for both long- and short-range expressions. The left end of the plots corresponds to vanishing functionals (i.e., $\mu=0$ for long-range functionals and $\mu \to \infty$ for short-range functionals). The right end of the plots corresponds to the usual (Coulombic) LDA case.

\begin{figure}[t]
\includegraphics[scale=0.7]{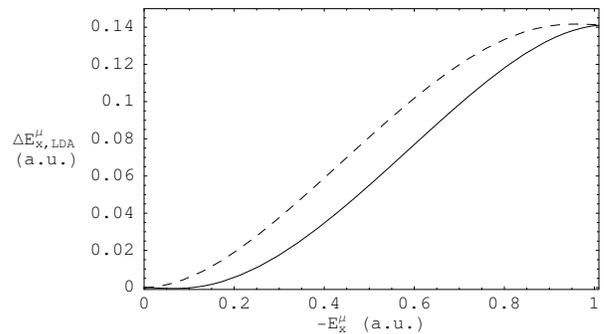}
\caption{LDA errors on $E_x^{\sr,\mu}$ (solid curve) and $E_x^{\lr,\mu}$ (dashed curve) with respect to $-E_x^{\sr,\mu}$ and $-E_x^{\lr,\mu}$ respectively, for the He atom.
}
\label{fig:DEx-he-erf}
\end{figure}

Let us consider first the LDA errors on $E_x^{\sr,\mu}$ and $E_x^{\lr,\mu}$, represented in Fig.~\ref{fig:DEx-he-erf}. For all values of the exchange energy (except of course at the end points), the LDA error on $E_x^{\sr,\mu}$ is systematically smaller than the error on $E_x^{\lr,\mu}$ (the difference is of the order of $0.025$ a.u. at an intermediate energy of $-0.5$ a.u.). Using the LDA for the short-range contribution to the exchange energy, rather than for the long-range contribution, allows, for a given energy, to do a smaller error on this energy, or equivalently, for a given error, to treat a larger part of the exchange energy. Note, in particular, that the LDA error on $E_x^{\sr,\mu}$ becomes vanishingly small toward the left end of the plot, corresponding to a very-short-range interaction. This is in agreement with the fact that the LDA becomes exact in the limit of a very short-range interaction~\cite{BurPerLan-PRL-94,GilAdaPop-MP-96,TouColSav-PRA-04}.

\begin{figure}[t]
\includegraphics[scale=0.7]{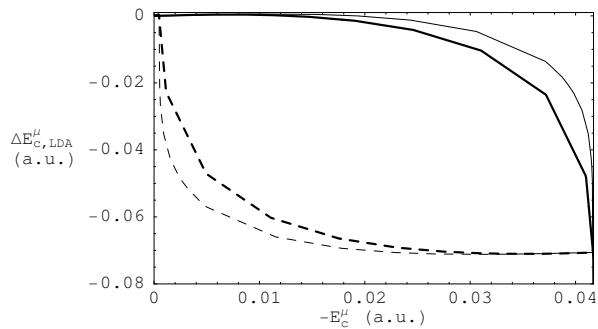}
\caption{LDA errors on $\bar{E}_c^{\sr,\mu}$ (thin solid curve), $E_c^{\sr,\mu}$ (thick solid curve), $E_c^{\lr,\mu}$ (thin dashed curve) and $\bar{E}_c^{\lr,\mu}$ (thick dashed curve) with respect to $-\bar{E}_c^{\sr,\mu}$, $-E_c^{\sr,\mu}$, $-E_c^{\lr,\mu}$ and $-\bar{E}_c^{\lr,\mu}$ respectively, for the He atom.
}
\label{fig:DEc-he-erf}
\end{figure}

Fig.~\ref{fig:DEc-he-erf} compares the LDA errors on $\bar{E}_c^{\sr,\mu}$, $E_c^{\sr,\mu}$, $E_c^{\lr,\mu}$ and $\bar{E}_c^{\lr,\mu}$. It clearly appears that, in the whole energy range (except at the end points), the LDA errors on the short-range contributions are much smaller (in absolute value) than those on the long-range contributions. This confirms the appropriateness of the LDA for short-range correlations, as often pointed out in the literature (see, e.g., Refs.~\onlinecite{LanPer-PRB-77,BurPer-IJQC-95,TouColSav-PRA-04}). Note that, contrary to the exchange case, the differences in the relative errors on the long-range and short-range contributions are important. For example, at an intermediate value of the energy of $-0.02$ a.u., the LDA errors on $E_c^{\lr,\mu}$ and $\bar{E}_c^{\lr,\mu}$ are $-0.0697$ a.u. and $-0.0673$ a.u., respectively, while the LDA errors on $E_c^{\sr,\mu}$ and $\bar{E}_c^{\sr,\mu}$ are as small as $-0.0017$ a.u. and $-0.0001$ a.u., respectively. One sees in addition that the LDA error on $\bar{E}_c^{\sr,\mu}$ is always significantly smaller (in absolute value) that the error on $E_c^{\sr,\mu}$. We conclude that the functional $\bar{E}_c^{\sr,\mu}$ is the best suited for a local density approximation.

\vspace{1.5cm}

\section{Conclusions}

As regards the long-range/short-range decomposition of the exchange functional of Eq.~(\ref{Exdecomp}), the results of this work suggest that the LDA is more accurate for the short-range contribution rather than for the long-range contribution. 

Concerning the long-range/short-range decomposition of the correlation functional according to Eq.~(\ref{Ecdecomp1}) or to Eq.~(\ref{Ecdecomp2}), this work confirms that the LDA is more accurate for the short-range contributions. The presented results suggest in addition that the short-range correlation functional appearing in the decomposition of Eq.~(\ref{Ecdecomp1}) is better suited for the LDA that the other short-range correlation functional appearing in the decomposition of Eq.~(\ref{Ecdecomp2}).

In the context of the long-range/short-range decomposition in DFT, the present paper gives clues about the part of the energy, better adapted to (local) density functional approximations. We hope that it will incite to more systematic studies of this topic.

\begin{acknowledgments}

It is our pleasure to dedicate this paper to Annick Goursot, for 
courageously defending DFT over the years, building the environment
which allowed us to pursue the present work.

We would also like to thank G. E. Scuseria for carefully reading our manuscript.

\end{acknowledgments}

%BIBLIOGRAPHY---------------------------------------------
\bibliographystyle{apsrev}
\bibliography{biblio}

\end{document}